\documentclass[12pt]{article}
\usepackage[utf8]{inputenc}
\usepackage{amsmath, amssymb, amsfonts,euscript,mathrsfs}
\usepackage{bm,bbm,latexsym}
\usepackage{graphicx}
\usepackage{rotating} 
\usepackage{epsfig, verbatim}
\usepackage{natbib}
\usepackage{times}
\usepackage{multirow}
\usepackage{xfrac}
\newcommand{\pkg}[1]{{\fontseries{b}\selectfont #1}} 
\usepackage{color}
\usepackage{colortbl,array} 
\usepackage{multirow,bigdelim}
\usepackage{booktabs}

\usepackage{setspace}

\def\linesparpage#1{
\baselineskip=\textheight
\divide\baselineskip by #1} 

\newcommand{\nc}{\newcommand}
\nc{\indep}{{\, \perp \! \! \! \perp  \,} }
\nc{\beq}{\begin{eqnarray*}}
\nc{\eeq}{\end{eqnarray*}}

\nc{\beqna}{\begin{eqnarray}}
\nc{\eeqna}{\end{eqnarray}}

\nc{\bct}{\begin{center}}
\nc{\ect}{\end{center}}

\nc{\bds}{\begin{description}}
\nc{\eds}{\end{description}}

\nc{\bit}{\begin{itemize}}
\nc{\eit}{\end{itemize}}

\nc{\bnu}{\begin{enumerate}}
\nc{\enu}{\end{enumerate}}

\nc{\bgt}{\begin{table}}
\nc{\bgtb}{\begin{center} \begin{tabular}}
\nc{\entb}{\end{tabular} \end{center} }
\nc{\ent}{\end{table}}

\title{\Large{\bf{Bayesian nonparametric generative models  for causal inference with missing at random covariates}}}

 \date{}
 
\begin{document}
 
 \maketitle


\author{\begin{center}Jason Roy$^{1,*}$ 
 Kirsten J. Lum$^1$, Michael J. Daniels$^2$, Bret Zeldow$^1$, Jordan Dworkin$^1$, and Vincent Lo Re III$^{1,3}$ \\ 
 $^1$Department of Biostatistics and Epidemiology, and Center for Causal Inference, University of Pennsylvania, Philadelphia, PA 19104, U.S.A.\\ 
$^2$Department of Statistics and Data Science, The University of Texas, Austin, TX 78712, U.S.A.  \\
$^3$Department of Medicine, Perelman School of Medicine, University of Pennsylvania, Philadelphia, PA 19104, U.S.A.\\
$^*$email: jaroy@upenn.edu\end{center}}

\bigskip

\begin{abstract}
We propose a general Bayesian nonparametric (BNP) approach to causal inference in the point treatment setting. The joint distribution of the observed data (outcome, treatment, and confounders) is modeled using an enriched Dirichlet process. The combination of the observed data model and causal assumptions allows us to identify any type of causal effect - differences, ratios, or quantile effects, either marginally or for subpopulations of interest. The proposed BNP model is well-suited for causal inference problems, as it does not require parametric assumptions about the distribution of confounders and naturally leads to a computationally efficient Gibbs sampling algorithm. By flexibly modeling the joint distribution, we are also able to impute (via data augmentation) values for missing covariates within the algorithm under an assumption of ignorable missingness, obviating the need to create separate imputed data sets. This approach for imputing the missing covariates has the additional advantage of guaranteeing congeniality between the imputation model and the analysis model, and because we use a BNP approach, parametric models are avoided for imputation. The performance of the method is assessed using simulation studies. The method is applied to data from a cohort study of human immunodeficiency virus/hepatitis C virus co-infected patients. 
\end{abstract}

%


\clearpage

\newcommand\independent{\protect\mathpalette{\protect\independenT}{\perp}}
\def\independenT#1#2{\mathrel{\rlap{$#1#2$}\mkern2mu{#1#2}}}

\makeatletter 
\newcommand*{\rom}[1]{\expandafter\@slowromancap\romannumeral #1@} 
\makeatother





\linesparpage{25}

\section{Introduction}
Bayesian methods have not been widely used for causal inference in observational studies. A possible reason for this is that causal inference in a likelihood-based framework often requires modeling the joint distribution of all of the observed data, including covariates or at least complex relationships between outcomes and confounders. For example, to estimate marginal causal effects, integration over the distribution of confounders is required (\citealp{robi:2000}). In settings with time-dependent confounding, the g-formula requires simulating effects of past treatment on future values of covariates and can be used to estimate causal effects (\citealp{robi:1986,youn:cain:2011,west:cole:2012}).   Because the dimension of covariates that need to be controlled for might be high, modeling the joint distribution of these covariates offers many opportunities for model misspecification. As a result, semiparametric methods that do not require specification of the joint distribution of the covariates have dominated the causal inference literature~\citep*[e.g.,][]{robi:hern:2000,vand:robi:2003,vand:2010a,vand:2010b,neug:fire:2013}.

However, recent developments in Bayesian nonparametric (BNP) modeling, along with increasing computing capacity, have opened the door to new, potentially powerful approaches to causal inference.  
In the point treatment setting, one option is to directly model the conditional distribution of the outcome given covariates using a dependent Dirichlet process \citep{mace:1999}. Modeling a conditional distribution directly is what  \citet{shah:neal:2009} refer to as discriminative models. Marginal causal effects can  be obtained by integrating the conditional distribution over the empirical distribution of the covariates. This approach was used by \citet*{roy:lum:2016} to directly parameterize causal effects from a marginal structural model. However, dependent Dirichlet process models can be computationally expensive. In addition, they do not easily allow for imputation of covariates within the model. 

An alternative to discriminative models is generative models, which model the joint distribution of the data (i.e., the outcomes and covariates). BNP generative models can be used to induce the conditional distribution of the outcome given covariates~\citep*{Mull:Erka:West:1996} and were used by~\citet*{xu:dani:2017} for causal inference using the propensity score. 
In this paper, we use a Dirichlet process model similar to that proposed by \citet{shah:neal:2009}. These models can easily accommodate discrete and continuous covariates as well as missing covariates under a specific assumption about the missingness. Perhaps more importantly,  they  can handle large $n$ and $p$ due to the local independence specification (described in Section~\ref{s:model}). We use a refinement to the model proposed by \citet{wade:duns:2014} to obtain a flexible, yet computationally tractable regression model for the outcome.   

The general approach to causal inference advocated in this paper can be briefly summarized as follows: first, specify the causal effects of interest and causal identification assumptions; second, model the joint distribution of the observed data using flexible BNP models; third, use post-processing steps (g-formula) to obtain estimates of causal effects. It is important to note here that modeling the observed data is a distinct step from computing causal effects. As a result, the same BNP model applied to the same observed data could be used to extract a variety of causal effect parameters, including average treatment effects, quantile treatment effects, causal effect of treatment on treated, conditional treatment effects, and so on.  This is the same approach that was taken by  \citet{dani:roy:2012} and \citet{kim:dani:2016} in causal mediation settings.

Modeling the full observed data distribution instead of just the conditional distribution of the outcome has many potential benefits, including efficiency gains, full posterior inference rather than just point estimates and confidence intervals, automatic imputation of missing data under an assumption of ignorable missingness, and a general way to account for uncertainty about a variety of assumptions.

The paper is organized as follows. Section 2 specifies causal identifying assumptions and causal effects that may be of interest. In Section 3 we develop a flexible model for the joint distribution of the observed data. Computations are described in Section 4.  There are simulation studies in Section 5 which compare the proposed approach to several semiparametric alternatives. The BNP approach is applied to data from a study of human immunodeficiency virus/hepatitis C virus (HIV/HCV) co-infected patients in Section 6 followed by a discussion in Section 7.

\section{Causal effects}\label{s:causaleffects}

Suppose we are interested in causal effects of treatment $A$ on outcome $Y$. We assume that treatment is discrete/categorical and not continuous, taking one of $q$ possible values. For the $i^{th}$ subject $(i=1, \ldots, n)$, the treatment is represented by a vector of indicator variables $A_i=(A_{1,i},\ldots,A_{q-1,i})^T$, where $A_{t,i}$ is an indicator for treatment category $t$. Most typically, $A_i$ will just be a single variable indicating whether the subject received the new treatment. Denote by $L_i$ a $p\times 1$ set of pre-treatment variables. 

Our goal is to identify causal effects from the observed data $(Y,A,L)$. In this section assume the joint distribution of the observed data, $p(y,a,l)$ is known. That is, our goal here is to specify causal effects of interest and identification assumptions, given that  $p(y,a,l)$ is known. In this section it does not matter whether the joint distribution of the observed data is described with few (very parametric) or infinitely many (nonparametric) parameters. Estimation of the joint distribution is a {\em distinct} step from defining causal effects and making identifying assumptions.
%
%

We consider definitions of causal effects that are functions of potential outcomes. Each subject has $q$ potential outcomes, $\{Y^a: a=0,\ldots,q-1\}$, where $Y^a$ is the outcome that would be observed if treatment was set to $a$. There are many possible causal effects that could be of interest to researchers. For simplicity, we will focus here on the situation where $q=2$. Some examples include:
\bit
\item $E(Y^1-Y^0$): average causal effect (continuous outcome) or average causal risk difference (binary outcome)
\item $E(Y^1)/E(Y^0)$: average causal relative risk (binary outcome) or average causal rate ratio (count outcome)
\item $E(Y^1-Y^0|V)$: conditional average causal effect (where $V\subset L$)
\item $E(Y^1-Y^0|A=1)$: average effect of treatment on treated
\item $F^{-1}_1(p)-F^{-1}_0(p)$, where $F^{-1}_a(p)$ is the $p$th quantile of the cumulative distribution function $P(Y^a\le y)$: a quantile causal effect (\citealp{xu:dani:2017}).
\eit

All of the above causal effects are functionals of the distribution of the potential outcomes. 
We can identify these causal effects from the following three assumptions.
The first assumption is consistency, which states that $Y^a=Y$ among subjects with $A=a$, for all $a$. That is, the potential outcome if we were to set $A=a$ is the same as the  outcome that we observe if $A=a$.  We next assume positivity $p(A=a|L)>0$ if $p(L)>0$. This implies that at each possible level of the confounders, each treatment level has non-zero probability. Finally, we assume ignorability, or $\{Y^a \indep A|L\}$. In other words, given confounders $L$, treatment can be thought of as randomly assigned.

These three assumptions imply
$F(y|A=a,L)=F(y^a|A=a,L)=F(y^a|L).$
We can therefore identify any functional of $F(y^a | L)$ from $p(Y,A,L)$:
\begin{align*}
E(Y^a)&=E\{E(Y|A=a,L)\} \\ 
E(Y^a|V=v)&=E\{E(Y|A=a,V=v,W)|V=v\},\\ 
E(Y^a|A=a')&=E\{E(Y|A=a,L)|A=a'\}=\int E(Y|A=a,L)dF(L|A=a'), \\
P(Y^a\le y)&=\int^y_{-\infty}\int p(Y|A=a,L)dF(L),
\end{align*}
where $L=(V,W)$. For $E(Y^a|A=a')$, integration is over $p(L|A=a')$, which is known if $p(Y,A,L)$ is known.

\section{BNP model for observed data}\label{s:model}

In order to estimate causal effects described in the Section~\ref{s:causaleffects}, we first need to estimate the joint distribution $p(Y,A,L)$. Let $X_i=(A_i^T,L_i^T)^T$ and consider estimation of $p(Y,X)$. While estimation of this joint distribution could be parametric or nonparametric, we propose a Bayesian nonparametric approach. This will allow us to flexibly model the joint distribution (whose parameter values are not of interest) while allowing ignorable missingness in  $L$ (more on the latter in Section~\ref{s:computations}). 

We propose to model the joint distribution of $(Y,X)$ using the following enriched Dirichlet process (EDP) mixture \citep{wade:mong:2011,wade:duns:2014}:
\begin{equation}
\begin{aligned}
\label{eq:edp}
Y_i|X_i, \theta_i &\sim p(y|x,\theta_i) \\
X_{i,r}|\omega_i &\sim p(x_r|\omega_i), \ \ \ \  r=1,\cdots,p+q-1\\
(\theta_i, \omega_i)|P& \sim P \\
P&\sim EDP(\alpha_{\theta},\alpha_\omega,P_0) .
\end{aligned}
\end{equation}
The notation $P\sim EDP(\alpha_{\theta},\alpha_\omega,P_0)$ means that $P_\theta \sim DP(\alpha_\theta, P_{0,\theta})$ and $P_{\omega|\theta}\sim DP(\alpha_\omega, P_{0,\omega|\theta})$ with base measures $P_0=P_{0,\theta}\times P_{0,\omega|\theta}$.   

This formulation implies that each subject $i$ has their own parameters $\theta_i$ and $\omega_i$. However, because $P$ is discrete \citep{ferguson1973bayesian}, some clusters of subjects will have the same $\theta_i$ and $\omega_i$. The number of clusters depends on the concentration parameters $\alpha_\theta$ and $\alpha_\omega$, where low values indicate fewer clusters. Typically DP models have a single concentration parameter. The {\em enrichment} of the usual DP is to have nested concentration parameters. This allows for more $x$-clusters than $y$-clusters, which is important because the dimension of $x$ will typically be much larger than that of $y$. Importantly, this is accomplished while keeping cluster membership dependent on both $y|x$ and $x$ through the nesting of the random partition. 
%
%
%
%

We assume a local generalized linear model for $p(y|x,\theta_i)$~\citep*{hann:blei:2011}. That is, 
\begin{align*}
p(y|x,\theta_i)=\exp\left\{\frac{Y_i\eta_i-b(\eta_i)}{a(\phi_i)}+c(y_i,\phi_i)\right\}
\end{align*}
where $g\{b'(\eta_i)\}=X_i\beta_i$ and $g\{\}$ is a link function. For example, if $Y$ is binary then
\[
Y_i|X_i, \theta_i \sim {\rm{Bern}}\{{\rm{logit}}^{-1}( X\beta_i)\}
\]
where $\theta_i=\beta_i$  and $X$ is the design matrix involving $A$ and $L$. In the linear regression case, $\theta_i$ would include both regression coefficients and a variance. 

An important aspect of this model is that covariates  $\bf X$ are assumed to be locally independent. That is given $\omega_i$, covariates are  independent. Two subjects in the same subcluster would have similar values of $X$. It is a well known property of random variables that dependence between them decreases as the window under consideration shrinks. As an illustration, consider bivariate normal random variables $x_1$ and $x_2$ with mean 0, variance 1, and correlation 0.9. In that case, $cor\{x_1,x_2|x_1\in (0,0.2),x_2\in (0,0.2)\}\approx 0.02$. The local independence assumption makes it easy to include many continuous and discrete confounders, because the joint distribution is just a product of marginal distributions. In addition, computations are considerably faster because covariance matrices for the joint distribution of confounders are not needed. Note that while we assume that locally the generalized linear model is correctly specified for $y$ and $x$ and that the $x$'s are independent from each other, globally all of the variables are dependent with potentially non-linear relationships. 
%
%

The EDP model (\ref{eq:edp}) can equivalently be represented with the square-breaking formulation \citep{wade:mong:2011}, which is a generalization of the standard stick-breaking representation of DP models \citep{Seth:1994}. The joint distribution of the observed data for subject $i$ can be written
\[
f(y_i,x_i|P)= \sum_{j=1}^\infty  \gamma_j  \sum_{l=1}^\infty\gamma_{l|j} K(y_i|x_i,\theta_{j})K(x_i|\omega_{l|j}),
\]
where $j$ indexes the $y$-clusters and the $K()$ are the kernels of the corresponding distributions. 
The weights have priors $\gamma_j'\sim$Beta$(1,\alpha_\theta)$ and $\gamma_{l|j}'\sim$Beta$(1,\alpha_\omega)$, where $\gamma_j=\gamma_j'\prod_{r<j}(1-\gamma_r')$ and  $\gamma_{l|j}=\gamma_{l|j}'\prod_{m<l}(1-\gamma_{m|j}')$.

The conditional distribution implied by the joint model is $p(y|x)=\sum_{j=1}^\infty w_j(x)K(y|x,\theta_j)$, where
\[
w_j(x)=\frac{\sum_{l=1}^\infty \gamma_{l|j}K(x|\omega_{l|j})} {\sum_{h=1}^\infty \gamma_h \sum_{l=1}^\infty \gamma_{l|h} K(x|\omega_{l|h})}.
\]
Notice that the weights $w_j(x)$ depend on $x$. Therefore, even though $K(y|x,\theta_j)$ is a generalized linear model, $p(y|x)$ is a computationally tractable, flexible, non-linear, non-additive model. 

\section{Computations}\label{s:computations}

We use a Gibbs sampler to obtain draws from the posterior distribution. In particular, we use an extension  of \citet{neal:2000} Algorithm 8 to accommodate nested clustering. This approach alternates between sampling cluster membership for each subject and sampling values of the parameters, given the cluster partitioning. Sampling cluster membership is not complex due to the closed form resulting from the P\'{o}lya urn in the collapsed Gibbs sampler. 
%
%

\begin{figure}
\begin{center}
 \includegraphics[width=5in]{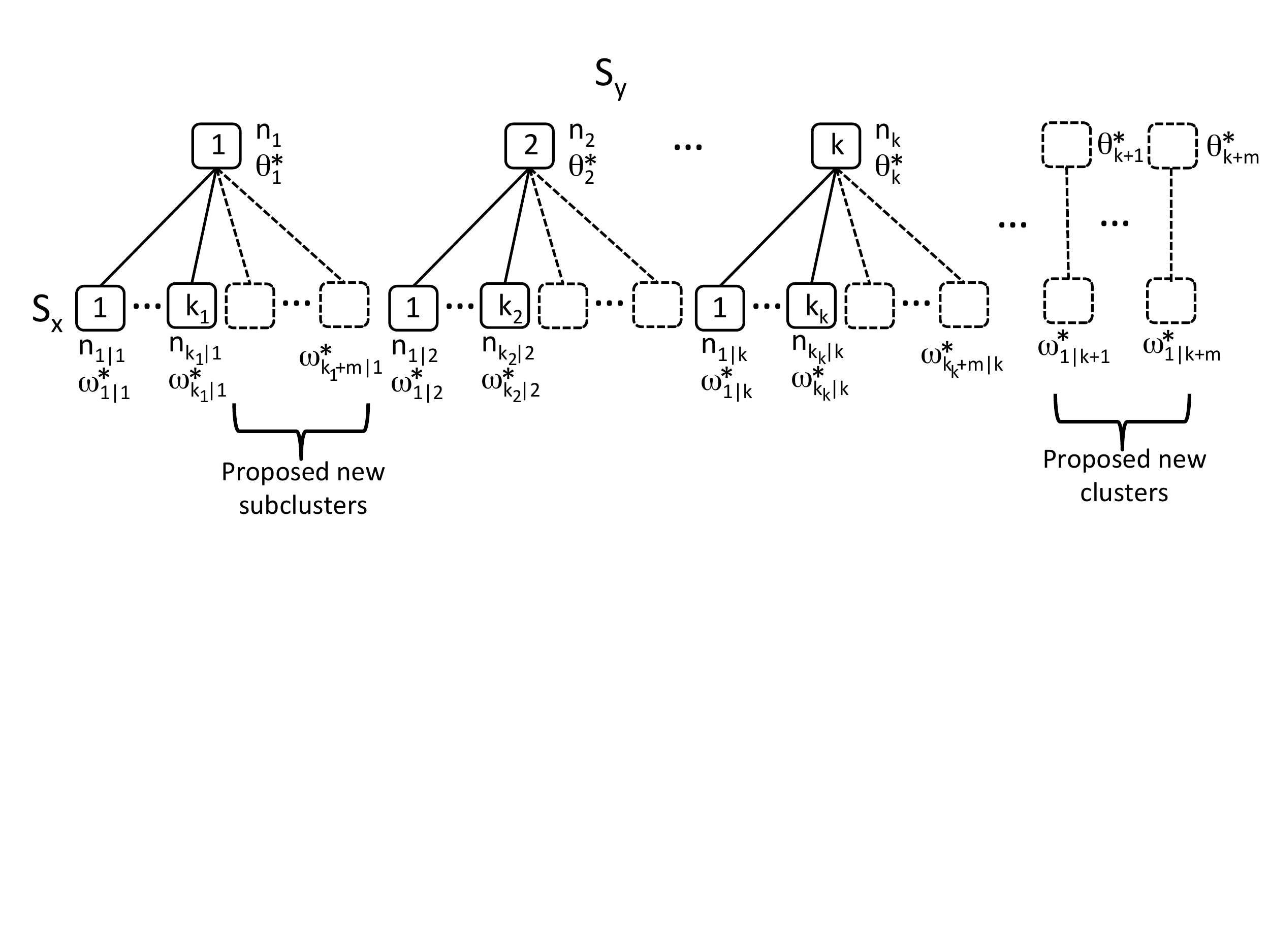}
\caption{Diagram of current clusters, subclusters, and values of parameters, along with proposed new clusters and subclusters. When updating cluster membership, the probability of being in each current and proposed new S is computed for subject i, and they are then randomly assigned to an S from the corresponding multinomial distribution. Once all subjects have been assigned to a cluster at a given iteration in the Gibbs sampler, then the parameters are updated, given cluster membership.}  
\label{fig:diagram}
\end{center}
\end{figure}

Here, we briefly describe the Gibbs sampling steps. Detailed steps are given in Appendix A. 
Following \citet{wade:duns:2014}, let $s_i=(s_{i,y},s_{i,x})$ denote cluster membership for subject $i$. Note that the value of $s_{i,x}$ is only meaningful in conjunction with $s_{i,y}$, as it describes which cluster within $s_{i,y}$ it belongs. The basic steps in the Gibbs sampler are as follows. 
We sample $s_i$ for each subject, and then, given $s$, we sample parameters $\theta$ and $\omega$ from their conditional distributions, given data. Denote by $\theta^*_j$ the $\theta$ that is associated with the $j^{th}$ currently non-empty $y$-cluster $(j=1,\ldots,k)$, with $\omega^*_{l|j}$ defined similarly and denote by $k_j$  the number of currently non-empty $x$-clusters within the $j^{th}$ $y$-cluster. The clusters, subclusters and their corresponding parameters are depicted in Figure~\ref{fig:diagram}. To obtain a new draw of $s_i$ given the current partition and current cluster-specific parameters, we simply draw from a multinomial distribution (see appendix for multinomial probabilities). Next, we provide further details on two additional components necessary for the causal inference setting, estimation of casual effects and imputation of missing covariates.  

{\bf{Post-processing steps for estimation of causal effects.}}
Once we have obtained draws of the parameters from the posterior distribution, we can compute any functionals of the distribution of potential outcomes.  Here we focus on the following expectations of the potential outcomes: 
$E(Y^a)$, $E(Y^a|V)$, or $E(Y^a|A=a')$.

Suppose, for example, that we would like to obtain draws from the posterior of $E(Y^a)$. Recall that $E(Y^a)=E\{E(Y|A=a,L)\}$
and we assume a GLM within clusters, with $E(y|a,l,\theta_j^*)=g^{-1}({\mathcal{X}}\beta_j^*)$, where ${\cal{X}}=(1,a^T,l^T)$. Thus, we can begin by obtaining a draw of $E(Y|A=a,L=l)$ for given values of covariates $l$. Given current values
of the parameters, $\{\theta^*, \omega^*,s\}$, from the Gibbs sampler, we can compute $E(Y|A=a,L=l,\theta^*,\omega^*,s)$ as

\[
E(Y|A=a,L=l,\theta^*,\omega^*,s)=\frac{w_{k+1}(a,l)E_0(y|a,l)+\sum_{j=1}^k w_j(a,l)E(y|a,l,\theta_j^*)}
{w_{k+1}(a,l)+\sum_{j=1}^k w_j(a,l)},
\]
where
\[
w_{k+1}(a,l)=\frac{\alpha_\theta}{\alpha_\theta+n}K_0(a,l),
\]
\[
w_j(a,l)=\frac{n_j}{\alpha_\theta+n}\left\{\frac{\alpha_\omega}{\alpha_\omega+n_j}K_0(a,l)+\sum_{l=1}^{k_j} 
\frac{n_{l|j}}{\alpha_\omega+n_j}K(a,l;\omega^*_{l|j})\right\}.
\] 
The terms $K_0(a,l)$ and $E_0(y|a,l)$ are the distribution and mean, respectively, after integrating the parameters over the prior distributions. That is, $K_0(a,l)=\int K(a,l;\omega)dF_0(\omega)$ and $E_0(y|a,l)=\int E(y|a,l,\theta)dF_0(\theta)$. For non-conjugate distributions, Monte Carlo (MC) integration can be used to obtain these quantities.

We can then obtain a draw from the marginal distribution $E(Y^a)$ by integrating over the distribution of $L$ using MC integration. For this, we must first draw $M$ samples from $p(L,s)$ as follows.\newline
For $m=1,\ldots,M$,
\begin{enumerate}
\item Draw $s^m_y$ from a multinomial $\{1,\cdots,k+1\}$ with probabilities $\left(\frac{n_1}{\alpha_\theta+n},\cdots,\frac{n_k}{\alpha_\theta+n},\frac{\alpha_\theta}{\alpha_\theta+n}\right)$. 
\item If $s^m_y < (k+1)$, draw $s^m_x$ from a multinomial $\{1,\cdots,k_j+1\}$ with probabilities $\left(\frac{n_{1|j}}{\alpha_\omega+n_j},\cdots,\frac{n_{k_{j}|j}}{\alpha_\omega+n_j},\frac{\alpha_\omega}{\alpha_\omega+n_j}\right)$; \newline
else, $s^m_x=1$.

\item Draw $L^m$ from $p(x|\omega^*_{{s^m_x}|{s^m_y}})$, where, if $s_y^m=(k+1)$ or if $s_x^m=(k_j+1)$ (i.e., if a new cluster is opened up), $\omega^*_{{s^m_x}|{s^m_y}}$ is drawn from the prior distribution.

\end{enumerate}

Once we have obtained $M$ values $(l^m,s^m)$, we can approximate the integral as follows:
\[
E(Y^a)\approx \frac{1}{M}\sum_{m=1}^M E(Y|A=a,L=l^m,\theta^*_{s^m_y},\omega^*_{{s^m_x}|{s^m_y}},s^m).
\]
Computing this separately for $a=1$ and $a=0$, for example, would allow us to obtain a draw of a causal effect, such as $E(Y^1)-E(Y^0)$. For causal effects conditional on $V=v$ or $A=a$, we essentially repeat the above steps, but integrate over the conditional distribution of $W|V=v$ or $L|A=a$, respectively, rather than the marginal distribution of $L$.

Because this is a post-processing step, its computation is not necessary for the Gibbs sampler used to sample the observed data model parameters. Therefore, to improve computational efficiency, draws of the causal effect parameters would not need to be obtained for every draw of the Gibbs sampler and could be done in parallel.

{\bf{Imputation of missing covariates.}} Missing values of covariates ($L$'s) can be dealt with using data augmentation under an assumption of ignorable missingness. Because we have already specified a full model for $(Y,A,L)$, we simply need to obtain draws of missing $L$'s from the appropriate conditional posterior distribution at each iteration in the Gibbs sampler. Suppose $L_{i,r}$ is a binary covariate that is missing for subject $i$. At each step in the Gibbs sampler, we do the following. Denote by $\omega_i^r$ the current value of binomial probability parameter for the $r$th covariate. Note that this value of $\omega$ is based on the cluster assigned to subject $i$. Denote by $X_i^{[k]}$ the vector $X_i$ in which covariate $L_{i,r}$ is set to a value of $k$. We draw $L_{i,r}$ from a binomial distribution with probability 
\[
\frac{\omega_i^r g^{-1}(X_i^{[1]}\beta_i)}
{\omega_i^r g^{-1}(X_i^{[1]}\beta_i)+(1-\omega_i^r)g^{-1}(X_i^{[0]}\beta_i)}.
\]

To draw values for missing continuous covariates, we use the Metropolis-Hastings algorithm with a random walk candidate distribution. The posterior for a missing continuous covariate $L_{i,r}$ is proportional to $K(L_{i,r}|\omega_i)g^{-1}(X_i\beta_i)$.


\section{Simulation studies}

We carry out simulations studies under four different data generating models to assess the performance of the proposed BNP method. In the first two simulation scenarios, there is a binary outcome and a binary treatment. The causal parameters of interest are a marginal relative risk, $\psi_{rr}=E(Y^1)/E(Y^0)$ and a marginal risk difference $\psi_{rd}=E(Y^1)-E(Y^0)$. In the last two scenarios, there is a continuous outcome and a binary treatment. The causal parameter of interest is the average causal effect $\psi=E(Y^1)-E(Y^0)$.

{\bf{Methods.}} In each simulation scenario, we estimate the causal parameter(s) using several methods and compare performance with that of the proposed BNP method. 
For the first two methods, inverse probability of treatment weighting (IPTW) and targeted maximum likelihood estimation (TMLE), we estimate the propensity score with a logistic model assuming an additive, linear form of the covariates, $L$. In scenarios 1-3, this propensity score model is correctly specified (i.e.~its functional form matches that of the model used to generate the treatment data and all of the necessary covariates are included).  In scenario 4, this propensity score model is misspecified because the treatment is generated using a complex functional form of covariates that in practice would most likely not be correctly specified. To estimate the outcome model in the TMLE approach, we  use Super Learner (\citealp*{vand:poll:2007}), which  is an ensemble machine learning method that uses cross-validation to weigh different prediction algorithms. We use three algorithms (glm, step, knn) for binary outcomes  (simulation scenarios 1 and 2) and three algorithms (glm, step, polymars) for continuous outcomes (scenarios 3 and 4), implemented using the R package \pkg{tmle} (\citealp*{grub:vand:2012}). Lastly, in scenarios 1 and 2 only, we compare the proposed BNP approach with a parametric Bayesian approach in which we fit fully Bayesian logistic regression models, with the treatment and covariates included in the model as additive, linear predictors. In scenario 2, the specified parametric distribution does not match the data generating distribution for the outcome. Average causal effects are obtained by averaging over the empirical distribution of the covariates.

For the proposed BNP approach, we first standardize the continuous covariates. We then assume
\begin{equation}
\begin{aligned}
\label{eqn:model}
P&\sim EDP(\alpha_\theta,\alpha_\omega) \\
(\beta_i, \pi_i, \mu_i,\tau^2_i, \sigma^2_i)|P&\sim P \\
X_{i,r}|\pi_{i}^r &\sim \rm{Bern}(\pi_{i}^r), \ r=1,\cdots,1+p_1 \\
X_{i,r}|\mu_{i}^r,\tau^{2,r}_{i}&\sim N(\mu_{i}^r,\tau_{i}^{2,r}), \  r=1+p_1+1,\cdots,1+p_1+p_2 \\
Y_i|X_i,\beta_i &\sim \rm{Bern}\{\rm{logit}^{-1}({\mathcal{X}}_i\beta_i)\} \ {\mbox{(scenarios 1 and 2)}} \\
Y_i|X_i,\beta_i &\sim N({\mathcal{X}}_i\beta_i, \sigma_i^2) \ {\mbox{(scenarios 3 and 4)}}
\end{aligned}
\end{equation}
where $\omega_i=(\pi_i,\mu_i,\tau^2_i)$ and ${\mathcal{X}}_i=(1,A_i,L_i^T)$. The prior for $\beta$ is $N(\beta_0, \tau_{\beta}^2I)$. We set $\beta_0$ to the maximum likelihood estimate from an ordinary logistic regression of $Y$ on ${\mathcal{X}}$ and $\tau^2_\beta=4$. That is, our prior guess is that cluster specific logistic regression coefficients will equal the corresponding coefficients from a logistic regression model applied to all of the data, with uncertainty in that guess reflected by the standard deviation of 2 (weakly informative). We assume conjugate priors $p_0(\pi^r)=$Beta$(a_{x},b_{x})$ for binary covariate parameters, where $a_{x}=b_{x}=1$ for $r=1,\ldots,p_1+1$. Priors for the continuous covariate parameters are $p_{0\omega}(\tau^{2,r})=$Scale Inv-$\chi^2(\nu_{0}, \tau^2_{0})$ and $p_{0\omega}(\mu^{r}|\tau^{2,r})=N(\mu_{0}, \tau^{2,r}/c_0)$,  where $\nu_{0}=2, \tau^2_{0}=1, c_0=0.5,$ and $\mu_{0}=0$, for $r=(1+p_1+1),\ldots,(1+p_1+p_2)$. For the concentration parameters $\alpha_\theta$ and $\alpha_\omega$, we assume Gam$(1,1)$ priors. For scenarios with $n=250$, we used a burn-in of 10,000 and then used an additional 90,000 Gibbs samples for posterior inference. For $n=1000$ or $n=3000$, we found that 1,000 draws from the Gibbs samplers were sufficient for a burn-in period and that 19,000 additional draws were enough to accurately capture the posterior. 

For each scenario and each method, we tested the methods on 1,000 generated datasets, and we report the absolute bias, empirical standard deviation (ESD), coverage probability, and the width of 95\% credible or confidence intervals. 

\subsection{Scenario 1: Binary outcome, simple functional forms}

We simulated a binary treatment, two binary covariates, two continuous covariates, and the binary outcome as follows:
\begin{align*}
L_1&\sim  {\rm{Bern}}(0.2), \\
L_2|L_1& \sim  {\rm{Bern}}\{{\rm{logit}}^{-1}(0.3+0.2L_1)\}, \\
L_3|L_1, L_2 &\sim N(L_1-L_2, 1^2), \\
L_4|L_1,L_2,L_3& \sim N(1+0.5L_1+0.2L_2-0.3L_3, 2^2), \\
A|L_1,\cdots,L_4 &\sim  {\rm{Bern}}\{{\rm{logit}}^{-1}(-0.4+L_1+L_2+L_3-0.4L_4)\}, \\
Y|A,L_1,\cdots,L4 &\sim  {\rm{Bern}}\{{\rm{logit}}^{-1}(-0.5+0.78A-0.5L_1-0.3L_2+0.5L_3-0.5L_4)\}.
\end{align*}

 The true causal parameters are $\psi_{rr}=1.5$ and $\psi_{rd}=0.13$. 

{\bf{Missing data.}} We carried out an additional analysis for the BNP approach only, where we randomly deleted values of $L$ based on the following probabilities:

$L_1$ missing with probability logit$^{-1}(-2+L_2+Y)$,

$L_2$ missing with probability logit$^{-1}(2+L_3+A)$,

$L_3$ missing with probability logit$^{-1}(-1.5-A+Y)$, and

$L_4$ missing with probability logit$^{-1}(-.9-L_1-L_2)$.

This results in about 20\% missing values for each covariate. We then analyzed the data using the data augmentation approach described in Section 4.

{\bf{Results.}} The results are given in Table~\ref{tab:scen1}. The proposed BNP approach performed well with nearly unbiased point estimates and the smallest ESDs, matching the performance of the correctly specified parametric Bayesian model. This is because the BNP model essentially settled on one cluster - making it equivalent to the parametric Bayesian model. Thus, in this simulation, no efficiency price was paid for fitting the more complex model. TMLE and IPTW also had low bias, but had slightly higher ESD than the Bayesian approaches.

The above results are all for the case of complete data. When fitted to data with missing covariates, the BNP approach had no bias or very little bias, similar in magnitude to the other approaches (fitted to complete data). The ESD increased slightly relative to the BNP approach when all covariates were observed (for example, 0.15 versus 0.16). Therefore, a benefit of the BNP approach is that it can be used with MAR covariates with little to no degradation of performance.

\subsection{Scenario 2: Binary outcome, mixture distribution}

For scenario 2, we generated a continuous confounder, binary treatment, and a binary outcome that depends on the confounder in a complex way as follows:
\begin{align*}
L&\sim N(4,2^2), \\
A|L &\sim {\rm{Bern}}\{{\rm{logit}}^{-1}(1.3-0.8L)\}, \\
Y|A,L &\sim (p) {\rm{Bern}}\{{\rm{logit}}^{-1}(-0.8-0.1L+A)\}+(1-p){\rm{Bern}}\{{\rm{logit}}^{-1}(-2+0.45L)\}, 
\end{align*}
where
\[
p=\frac{2\exp\{-2(L-4)^2\}}{2\exp\{-2(L-4)^2\}+2\exp\{-2(L-6)^2\}}.
\]
 The true causal parameters are $\psi_{rr}=1.4$ and $\psi_{rd}=0.155$. 

{\bf{Missing data.}} We carried out an additional analysis for the BNP approach only, where we randomly deleted values of $L$ with probability logit$^{-1}(-2+A+Y)$. This resulted in about 20\% missing values of $L$.

{\bf{Results.}} The results are given in Table~\ref{tab:scen2}.
Here, the outcome was generated using a mixture distribution; however, the Bayesian parametric method modeled the outcome using a logistic model with all covariates included as additive, linear terms. The parametric approach therefore performed poorly (with large bias and coverage dropping below 20\% in the $n=1000$ case). In contrast, the BNP approach had absolute bias of 0.04 or below at both sample sizes, good coverage, and a lower ESD than IPTW and TMLE. Deleting about 20\% of the covariate values, and then imputing within the BNP approach, resulted in credible intervals that were typically only about 2\% wider than in the complete data scenario using the BNP approach.

\subsection{Scenario 3: Continuous outcome, complex functional forms}

For scenario 3, we considered a continuous outcome. There were 4 confounders, 
$L_1,\ldots,L_4$, which were distributed as multivariate normal with mean 0, variance 1 and covariance 0.3. We then generated a binary treatment and continuous outcome as follows:
\[
A|L\sim {\rm{Bern}}\{{\rm{logit}}^{-1}(0.3\sum_{j=1}^4 L_j)\},
\]
\[
Y|A,L \sim p N(\mu_1,1)+(1-p) N(\mu_2,4^2),
\]
where
\begin{align*}
p&=\frac{\exp\{-2(L_1+1)^2\}}{\exp\{-2(L_1+1)^2\}+\exp\{-2(L_1-2)^2\}}, \\
\mu_1&=-4+2A-0.5L_2-L_3+0.5L_4,  \\
\mu_2&=4+0.4A+0.5L_2^2-0.8 L_3(L_3>0).
\end{align*}
The true average causal effect is $\psi=1.503$.

{\bf{Results.}} The results are given in Table~\ref{tab:scen3}.
In this scenario, both IPTW and TMLE were unbiased, with TMLE having the lower (of the two) ESD. The BNP approach had a very small bias that decreased as the sample size increased and it has the smallest ESDs. The BNP approach had a bit of undercoverage relative to the other approaches, but the comparison is not completely fair since the other approaches used a correctly specified propensity score (i.e., estimated using the same functional form and necessary covariates as in the treatment generating model).

\subsection{Scenario 4: Continuous outcome, complex treatment, many covariates} 

For scenario 4, we generated data similar to scenario 3; however, we added an additional 80 covariates for a total of 84 to test how well the methods handled a large number of covariates. Specifically we generated 40 binary covariates $(L_1,\ldots,L_{40})$, independently distributed as Bernoulli$(0.5)$, and 44 continuous covariates $(L_{41},\ldots,L_{84})$, distributed as multivariate normal with mean 0, variance 1, and correlation 0.3. We then generated a binary treatment in a complex way as follows:
\begin{align*}
A|L_{41},\ldots,L_{44}\sim \rm{Bern}\bigg[{\rm{logit}}^{-1}\Big\{(\lambda){\rm{logit}}^{-1}(0.6L_{41}L_{42}-0.2L_{43}^2)+(1-\lambda){\rm{logit}}^{-1}(0.7L_{41}-0.4L_{43}L_{44})\Big\}\bigg],
\end{align*}
where
\begin{align*}
\lambda &= \frac{\exp\{-2(L_{42}+1)^2\}}{\exp\{-2(L_{42}+1)^2\}+\exp\{-2(L_{42}-2)^2\}}.\\
\end{align*}

The outcome was generated as follows:
\[
Y|A,L_{41},\ldots,L_{44} \sim p N(\mu_{1},1)+(1-p) N(\mu_{2},4^2),
\]
where
\begin{align*}
p&=\frac{\exp\{-2(L_{41}+1)^2\}}{\exp\{-2(L_{41}+1)^2\}+\exp\{-2(L_{41}-2)^2\}}, \\
\mu_{1}&=-4+2A-0.5L_{42}-L_{43}+0.5L_{44},  \\
\mu_{2}&=4+0.4A+0.5L_{42}^2-0.8L_{43}(L_{43}>0).
\end{align*}
The true average causal effect is $\psi=1.503$. 

{\bf{Results.}} The results from scenario 4 are given in Table~\ref{tab:scen4mis}. In each of the methods, all 84 covariates were treated as confounders and included in all models. The BNP approach performed well with very small bias and ESDs only slightly larger than those of scenario 3. That is, adding unnecessary covariates seemed to mostly affect the uncertainty in the BNP estimates. For IPTW and TMLE, we estimated the propensity score using a logistic model assuming an additive, linear form of the 84 covariates. Since this model did not account for the complex form of the generating model for $A$, we observed an increase in bias relative to scenario 3. All of the methods showed some decrease in coverage for this scenario, especially IPTW for the n=3000 case.  

\section{Application}\label{s:application}

 Antiretroviral therapy (ART) is recommended for all human immunodeficiency virus (HIV) / chronic hepatitis C virus (HCV)-coinfected patients. ART regimens often include drugs from the nucleoside reverse transcriptase inhibitor (NRTI) class. There is concern that some drugs in the NRTI class (didanosine, stavudine, zidovudine, and zalcitabine) might cause depletion of mitochondrial DNA, leading to liver injury. We apply the proposed BNP approach to compare outcome $Y$ (death within 2 years) among those prescribed mitochondrial toxic NRTI (mtNRTI)-containing ART regimen to those prescribed other NRTI-containing ART regimen. 

To address this question, we used data from a study of HIV/HCV patients who newly initiated ART within the Veterans Aging Cohort Study (VACS) \citep{fult:skan:2006}. The study population included co-infected patients who newly initiated an ART-regimen that include NRTIs (either mtNRTIs or other NRTIs) from 2002 to 2009. There were a total of $n=1747$ patients included in the study. As can be seen from table S1 in Appendix B, use of mtNRTI-containing ART regimens as first-line therapy decreased over time, going from a large majority of cases in 2002 to a small minority of cases in 2009.

Our exposure $A$ was set to 1 for patients initiating an ART regimen that included an mtNRTI, and set to 0 for patients initiating an ART regiment that included some other NRTI. The outcome was all-cause mortality and we focused on the event occurring within 2 years of ART initiation. We had follow-up data on all patients through 2011, and so even patients who initiated ART in 2009 had 2 years of follow-up data. There were 76 deaths out of 836 patients in the mtNRTI group, and 89 deaths out of 911 patients in the other NRTI group. Our causal parameter of interest is the relative risk $\psi_{rr}=E(Y^{1})/E(Y^{0})$.

Variables that were included in the model as confounders ($L$) included the following baseline demographics and clinical variables: age at baseline (years), race/ethnicity, body mass index, diabetes mellitus, alcohol dependence/abuse, injection/non-injection drug abuse, year of ART initiation, and exposure to other antiretrovirals associated with hepatotoxicity (i.e., abacavir, nevirapine, saquinavir, tipranavir). In addition, the following baseline laboratory variables were included in $L$: CD4 count, HIV RNA, alanine aminotransferase (ALT), aspartate aminotransferase (AST), and fibrosis-4 (FIB-4) score. 
The percentage of  missing data for each variable is as follows: ALT 1.3\%, AST 2.5\%, CD4 1.8\%, FIB-4 3.1\%. The percentage of patients with at least one missing variable is 4.8\%.

For the observed data, we used the model specified in (\ref{eqn:model}) with a logistic regression model for the outcome. The prior for $\beta$ is $N(\beta_0, \tau_0^2I)$. We set $\beta_0$ to the maximum likelihood estimate from an ordinary logistic regression of $Y$ on ${\mathcal{X}}$ and $\tau^2_0=4$. The other prior distributions were the same as those specified in the simulation studies. 

{\bf{Results.}} We ran three chains of the Gibbs sampler, each with 20,500 iterations. The chains mixed well, with convergence appearing to be reached by iteration 500. The Gelman-Rubin convergence diagnostic was 1.04, providing further evidence of convergence. We calculated the average causal relative risk at every $100$th draw of the sampler after a burn-in of 500 for a total of 200 iterations. The number of $y$-clusters $k$ and $x$-subclusters $k_j$ varied from iteration to iteration and depended on the most recent posterior sample of $\alpha_\theta$ and $\alpha_\omega$, with bigger values leading to more clusters. The posterior median and 95\% credible intervals (CI) for $\alpha_\theta$ and $\alpha_\omega$ were $0.63 (0.21, 1.43)$ and $0.74 (0.43, 1.16)$, respectively.  The value of $k$ tended to be about 4, while $k_j$ tended to range from 1 to 7. For example, at the last iteration of the first chain, there were $k=5$ y-clusters, with the following sample sizes in each subcluster: cluster $s_y=1$, $(36, 164, 134, 45, 32, 38, 76, 1)$; cluster $s_y=2$, $(171, 211, 131, 68, 18, 1)$;  cluster $s_y=3$, $(171, 281, 172, 50, 28)$;  cluster $s_y=4$, $(137, 30, 2, 1)$;   cluster $s_y=5$, (2). 

The posterior median and 95\% CI of the average causal relative risk (RR), $\psi_{rr}$, were $1.16 (0.87, 1.54)$; thus, there was a 16\% increased risk of death within 2 years comparing mtNRTI-containing ART regimens with other NRTI-containing ART regimens. However, the uncertainty about $\psi_{rr}$  is substantial enough that we cannot rule a small reduced risk of death (e.g., RR about 0.9) or a larger increased risk (e.g., RR about 1.5). The posterior distribution of $\psi_{rr}$ is displayed in Figure S1 of Web Appendix B, along with the trace plot.

\section{Discussion}

In this paper, we developed a fully Bayesian approach for causal inference that can handle discrete or continuous outcomes and categorical treatment. While the full distribution of outcome, treatment, and confounders is modeled, the proposed BNP approach allows for flexible modeling of these distributions, estimation of any functionals of the potential outcome distribution, and high-dimensional confounding. In addition, because we have a model for the joint distribution, imputation of missing covariates under ignorable missingness is straightforward and does not require multiply imputed datasets. 

Our simulations showed overall good performance of the BNP approach. It is worth noting that we found (in simulation 1) that if the base model of BNP is the true model, no efficiency price was paid for fitting the more complex model. Compared to IPTW and TMLE, the BNP approach had the smallest ESD for all scenarios and sample sizes. In scenario 4, we generated a complex treatment for which the propensity score was misspecified as it would be unlikely for the form of the true propensity model to be implemented in the semiparametric approaches. For this scenario, the BNP approach had the smallest bias and ESD. This scenario also had a relatively large number of non-confounders and each of the methods displayed some amount of undercoverage. Thus, future work on settings with many covariates that are not actually confounders is of interest. For example, zero-inflated or shrinkage priors for the coefficients in the BNP model could be explored. For the TMLE method, bootstrapping as opposed to using asymptotic confidence intervals, may improve the coverage.

In the BNP model for the observed data, we included a model for treatment; however, if the sample size in each treatment category is sufficient, one could alternatively condition on treatment and use a separate BNP model of the outcome and covariates for each treatment category. Also, the general EDP approach allows for $\alpha_\omega$ to be a function of $\theta$ (cf. eq. (1)). In our analyses we only included a single $\alpha_\omega$ parameter. Thus, more complex models could be considered. An area for future research is the extension to the time-varying confounding setting on which we are currently working.


\section*{Acknowledgments}
The work was partially supported by NIH grants CA183854 and R01 GM112327.

\clearpage

\begin{table}
\caption{Results from simulation scenario 1: binary outcome, simple functional forms. The true values were $\psi_{rr}=1.5$ and $\psi_{rd}=0.13$. IPTW uses a correctly specified propensity score. TMLE uses a correctly specified propensity score and Super Learner with 3 prediction algorithms for the outcome model. The Bayesian parametric (Bayesian par.) approach uses a correctly specified logistic regression model and integrates over confounders using the empirical distribution. BNP is the proposed method. `BNP missing data' is the BNP approach, with data augmentation, applied to a data set where approximately 20\% of the covariate values were set to missing. The first four methods were applied to the full data set with no missing covariate values. Bias is the absolute bias and ESD is the empirical standard deviation. Results are from 1000 simulated datasets.}
\label{tab:scen1}
\begin{center}
\begin{tabular}{lrccccccc}
\hline
\hline
 & \multicolumn{4}{c}{Relative risk, $\psi_{rr}$} & \multicolumn{4}{c}{Risk difference, $\psi_{rd}$} \\
Method &  Bias & Coverage & ESD & CI width  &  Bias & Coverage & ESD & CI width\\
\hline
& \multicolumn{8}{c}{$n=250$} \\
IPTW                &        0.09        &        0.96         &        0.43&  1.76 &0.00 &0.96 &0.08 &0.33 \\
TMLE &        0.06        &        0.92         &        0.37&  1.31 &0.00 &0.91 &0.07 &0.25 \\
Bayesian par.                  &          0.05  &         0.93        & 0.33  &1.29 &0.00 &0.94 &0.06 &0.24           \\
 BNP                &  0.03       &    0.93            & 0.32 &1.20       & 0.00 & 0.93 & 0.06 & 0.23  \\ [.5ex]
BNP missing data &    0.05    &        0.94        & 0.33  & 1.31    & 0.00 & 0.94  & 0.07  & 0.25   \\
\hline
& \multicolumn{8}{c}{$n=1000$} \\
IPTW       &0.03	&0.97	&0.20	&0.84	&0.00	&0.97	&0.04	&0.17	  \\     
TMLE &0.01	&0.93	&0.18	&0.65	&0.00	&0.93	&0.04	&0.13	  \\     
 Bayesian par.                  & 0.01&	0.95&	0.15&	0.59&	0.00&	0.94	&0.03&	0.12	  \\     
 BNP              &   0.01       & 0.94 & 0.15                         & 0.58     &0.00 &0.94 &0.03 &0.12    \\ [.5ex]
BNP missing data  &      0.01       & 0.93 & 0.16                       & 0.63     &0.00 &0.93 &0.03 &0.13    \\ 
                                                        
\hline          
\end{tabular}
\end{center}
\end{table}

\begin{table}
\caption{Results from simulation scenario 2: binary outcome, mixture distribution. The true values were $\psi_{rr}=1.4$ and $\psi_{rd}=0.155$. IPTW uses a correctly specified propensity score. TMLE uses a correctly specified propensity score and Super Learner with 3 prediction algorithms for the outcome model. The Bayesian parametric (Bayesian par.) approach uses a misspecified logistic regression model. BNP is the proposed method. `BNP missing data' is the BNP approach, with data augmentation, applied to a data set where approximately 20\% of the covariate values were set to missing. The first four methods were applied to the full data set with no missing covariate values. Results are from 1000 simulated datasets.}
\label{tab:scen2}
\begin{center}
\begin{tabular}{lrccccccc}
\hline
\hline
 & \multicolumn{4}{c}{Relative risk, $\psi_{rr}$} & \multicolumn{4}{c}{Risk difference, $\psi_{rd}$} \\
Method &  Bias & Coverage & ESD & CI width  &  Bias & Coverage & ESD & CI width\\
\hline
& \multicolumn{8}{c}{$n=250$} \\
IPTW                & 0.02         &0.92  & 0.27                         &  1.27    & 0.01& 0.89&0.13 &0.44    \\     
TMLE    & 0.03       &0.86  & 0.32                         &  1.03   &0.02 &0.83 & 0.12& 0.36   \\   
Bayesian par.         &   0.36       & 0.65 & 0.25                         & 0.94     &0.12 &0.62 &0.08 &0.29    \\        
 BNP                &   0.04     &     0.93           & 0.26  & 0.97     &0.01 &0.93 &0.09 &0.34  \\ [.5ex]
BNP missing data    & 0.07      &    0.95            & 0.26  &  1.00   & 0.02 & 0.94 & 0.09 & 0.35 \\ 
\hline
& \multicolumn{8}{c}{$n=1000$} \\
IPTW  &  0.00 &	0.92 &	0.19 &	0.71 &	0.00	&0.92	&0.07&	0.26	  \\      
TMLE    &   0.02     &0.91  & 0.16                         &0.56     &0.01 &0.90 &0.06 & 0.20   \\     
 Bayesian par.     &      0.33	&0.19&	0.13	&0.46	&0.12&	0.17&	0.04	&0.14   \\        
 BNP              &  0.04      & 0.95  &  0.13  & 0.54     &0.02 &0.94 &0.05  & 0.20    \\    [.5ex]                                       
BNP   missing data      &  0.02     & 0.94  &  0.15  & 0.58     &0.01 &0.94 &0.05  & 0.21     \\    [.5ex]   
\hline          
\end{tabular}
\end{center}
\end{table}

\begin{table}[h]
\caption{Results from simulation scenario 3: continuous outcome, complex functional forms. The true average causal effect was $\psi=1.503$. IPTW and TMLE both use a correctly specified propensity score. TMLE uses Super Learner with 3 prediction algorithms for the outcome model. BNP is the proposed method. Results are from 1000 simulated datasets.}
\label{tab:scen3}
\begin{center}
\begin{tabular}{lrccc}
\hline
\hline
Method &  Bias & Coverage & ESD & CI width  \\
\hline
& \multicolumn{4}{c}{$n=1000$} \\
IPTW	  &0.00  &0.98		& 0.24&	1.18  \\    
TMLE     &0.00  &0.94		& 0.21&0.79	  \\    
 BNP      	  &0.09  &0.91		&0.19 &0.71		  \\  
							\hline

& \multicolumn{4}{c}{$n=3000$} \\
IPTW	  &0.00  &0.99		& 0.17&	0.68  \\    
TMLE    &0.00  &0.94		& 0.12&0.47	  \\       
 BNP      	  & 0.05 &	0.93	&0.10 &0.41		  \\  
\hline	  
\end{tabular}
\end{center}
\end{table}

\begin{table}[h]
\caption{Results from simulation scenario 4: continuous outcome, complex treatment, many covariates. The true average causal effect was $\psi=1.503$. IPTW and TMLE use a misspecified propensity score. BNP is the proposed approach. Results are from 1000 simulated datasets.}
\label{tab:scen4mis}
\begin{center}
\begin{tabular}{lrccc}
\hline
\hline
Method &  Bias & Coverage & ESD & CI width  \\
\hline
& \multicolumn{4}{c}{$n=1000$} \\
IPTW	  &0.26  &0.90		& 0.23&	1.10  \\    
TMLE       &0.11  &0.89		& 0.20&	0.76  \\     
 BNP      	& 0.05 & 0.87 & 0.23 & 0.72  \\  
							\hline

& \multicolumn{4}{c}{$n=3000$} \\
IPTW	    &0.27  &0.61		& 0.13&	0.61  \\     
TMLE        &0.09  &0.83		& 0.13&	0.43  \\        
 BNP      	& 0.06 & 0.89 & 0.11 & 0.41	  \\  
\hline	  
\end{tabular}
\end{center}
\end{table}

\clearpage


\begin{thebibliography}{}

\bibitem[\protect\citename{Daniels {\em et~al.}, }2012]{dani:roy:2012}
Daniels, M.~J., Roy, J.~A., Kim, C., Hogan, J.~W., and Perri, M.~G. 2012.
\newblock Bayesian inference for the causal effect of mediation.
\newblock {\em Biometrics}, {\bf 68}, 1028--1036.

\bibitem[\protect\citename{Escobar and West, }1995]{esco:west:1995}
Escobar, M.~D. and West, M. (1995). 
\newblock Bayesian density estimation and inference using mixtures.
\newblock {\em Journal of the American Statistical Association}, {\bf 90}, 557--588.

\bibitem[\protect\citename{Ferguson, }1973]{ferguson1973bayesian}
Ferguson, T.~S. 1973.
\newblock A Bayesian analysis of some nonparametric problems.
\newblock {\em The Annals of Statistics}, {\bf 1},  209--230.

\bibitem[\protect\citename{Fultz {\em et~al.}, }2006]{fult:skan:2006}
Fultz, S.L., Skanderson, M., Mole, L.A., Gandhi, N., Bryant, K., Crystal, S.,
  and Justice, A.C. 2006.
\newblock Development and verification of a "virtual" cohort using the National
  VA Health Information System.
\newblock {\em Medical Care}, {\bf 44(8 Suppl 2)}, S25--30.

\bibitem[\protect\citename{Gruber and van~der Laan, }2012]{grub:vand:2012}
Gruber, S., and van~der Laan, M.J. 2012.
\newblock tmle: An {R} package for targeted maximum likelihood estimation.
\newblock {\em Journal of Statistical Software}, {\bf 51}, 1--35.

\bibitem[\protect\citename{Hannah{, Blei, and Powell}, }2011]{hann:blei:2011}
Hannah, L.~A., Blei, D.~M., and Powell, W.~B. 2011.
\newblock Dirichlet process mixtures of generalized linear models.
\newblock {\em Journal of Machine Learning Research}, {\bf 12}, 1923--1953.

\bibitem[\protect\citename{Kim {\em et~al.}, }2016]{kim:dani:2016}
Kim, C., Daniels, M.J., Marcus, B.H., and Roy, J.A. 2016.
\newblock {A framework for Bayesian nonparametric inference for causal effects
  of mediation}.
\newblock {\em Biometrics},  doi: 10.1111/biom.12575.

\bibitem[\protect\citename{MacEachern, }1999]{mace:1999}
MacEachern, S.~N. 1999.
\newblock Dependent nonparametric processes.
\newblock {\em ASA Proceedings of the Section on Bayesian Statistical Science},
   50--55.

\bibitem[\protect\citename{M{\"u}ller{, Erkanli, and West}, }1996]{Mull:Erka:West:1996}
M{\"u}ller, P., Erkanli, A., and West, M. 1996.
\newblock Bayesian curve fitting using multivariate normal mixtures.
\newblock {\em Biometrika}, {\bf 83}, 67--79.

\bibitem[\protect\citename{Neal, }2000]{neal:2000}
Neal, R.~M. 2000.
\newblock Markov chain sampling methods for Dirichlet process mixture models.
\newblock {\em Journal of Computational and Graphical Statistics}, {\bf 9},
  249--265.

\bibitem[\protect\citename{Neugebauer {\em et~al.}, }2013]{neug:fire:2013}
Neugebauer, R., Fireman, B., Roy, J.~A., Raebel, M.~A., Nichols, G.A., and
  O'Connor, P.J. 2013.
\newblock Super learning to hedge against incorrect inference from arbitrary
  parametric assumptions in marginal structural modeling.
\newblock {\em Journal of Clinical Epidemiology}, {\bf 66}, S99--109.

\bibitem[\protect\citename{Robins, }1986]{robi:1986}
Robins, J.~M. 1986.
\newblock A new approach to causal inference in mortality studies with a
  sustained exposure period -- application to control of the healthy worker
  survivor effect.
\newblock {\em Mathematical Modeling}, {\bf 7}, 1393--1512.

\bibitem[\protect\citename{Robins, }2000]{robi:2000}
Robins, J.~M. 2000.
\newblock Marginal structural models versus structural nested models as tools
  for causal inference.
\newblock {\em Pages  95--133 of:} Halloran, M.~E., and Berry, D. (eds), {\em
  Statistical Models in Epidemiology, the Environment, and Clinical Trials}.
\newblock The IMA Volumes in Mathematics and its Applications, vol. 116.
\newblock Springer New York.

\bibitem[\protect\citename{Robins{, Hern\'{a}n, and Brumback}, }2000]{robi:hern:2000}
Robins, J.~M., Hern\'{a}n, M.~A., and Brumback, B. 2000.
\newblock {{M}arginal structural models and causal inference in epidemiology}.
\newblock {\em Epidemiology}, {\bf 11}, 550--560.

\bibitem[\protect\citename{Roy{, Lum, and Daniels}, }2017]{roy:lum:2016}
Roy, J., Lum, K.~J., and Daniels, M.~J. 2017.
\newblock A Bayesian nonparametric approach to marginal structural models for
  point treatments and a continuous or survival outcome.
\newblock {\em Biostatistics}, {\bf 18}, 32--47.

\bibitem[\protect\citename{Sethuraman, }1994]{Seth:1994}
Sethuraman, J. 1994.
\newblock A constructive definition of Dirichlet priors.
\newblock {\em Statistica Sinica}, {\bf 4}, 639--650.

\bibitem[\protect\citename{Shahbaba and Neal, }2009]{shah:neal:2009}
Shahbaba, B., and Neal, R. 2009.
\newblock Nonlinear models using Dirichlet process mixtures.
\newblock {\em Journal of Machine Learning Research}, {\bf 10}, 1829--1850.

\bibitem[\protect\citename{van~der Laan, }2010a]{vand:2010a}
van~der Laan, M.~J. 2010a.
\newblock {{T}argeted maximum likelihood based causal inference: {P}art {I}}.
\newblock {\em The International Journal of Biostatistics}, {\bf 6}, Article
  2.

\bibitem[\protect\citename{van~der Laan, }2010b]{vand:2010b}
van~der Laan, M.~J. 2010b.
\newblock {{T}argeted maximum likelihood based causal inference: {P}art
  {I}{I}}.
\newblock {\em The International Journal of Biostatistics}, {\bf 6}, Article
  3.

\bibitem[\protect\citename{van~der Laan and Robins, }2003]{vand:robi:2003}
van~der Laan, M.J., and Robins, J.M. 2003.
\newblock {\em {Unified Methods for Censored Longitudinal Data and Causality}}.
\newblock Springer Series in Statistics.
\newblock New York, New York: Springer.

\bibitem[\protect\citename{van~der Laan{, Polley, and Hubbard}, }2007]{vand:poll:2007}
van~der Laan, M.J., Polley, E.C., and Hubbard, A.E. 2007.
\newblock Super {L}earner.
\newblock {\em Statistical Applications in Genetics and Molecular Biology},
  {\bf 6}, 1--21.

\bibitem[\protect\citename{Wade{, Mongelluzzo, and Petrone}, }2011]{wade:mong:2011}
Wade, S., Mongelluzzo, S., and Petrone, S. 2011.
\newblock An enriched conjugate prior for Bayesian nonparametric inference.
\newblock {\em Bayesian Analysis}, {\bf 6}, 359--385.

\bibitem[\protect\citename{Wade {\em et~al.}, }2014]{wade:duns:2014}
Wade, S., Dunson, D.~B., Petrone, S., and Trippa, L. 2014.
\newblock Improving prediction from Dirichlet process mixtures via enrichment.
\newblock {\em Journal of Machine Learning Research}, {\bf 15}, 1041--1071.

\bibitem[\protect\citename{Westreich {\em et~al.}, }2012]{west:cole:2012}
Westreich, D., Cole, S.~R., Young, J.~G., Palella, F., Tien, P.~C., Kingsley,
  L., Gange, S.~J., and Hernan, M.~A. 2012.
\newblock {{T}he parametric g-formula to estimate the effect of highly active
  antiretroviral therapy on incident {A}{I}{D}{S} or death}.
\newblock {\em Statistics in Medicine}, {\bf 31}, 2000--2009.

\bibitem[\protect\citename{Xu{, Daniels, and Winterstein}, }2017]{xu:dani:2017}
Xu, D., Daniels, M.J., and Winterstein, A.G. 2017.
\newblock Causal inference on quantiles with application to electronic health
  records.
\newblock {\em submitted}.

\bibitem[\protect\citename{Young {\em et~al.}, }2011]{youn:cain:2011}
Young, J.G., Cain, L.E., Robins, J.M., O'Reilly, E.J., and M.A., Hern{\'a}n.
  2011.
\newblock Comparative effectiveness of dynamic treatment regimes: an
  application of the parametric g-formula.
\newblock {\em Statistics in Biosciences}, {\bf 3}, 119--143.

\end{thebibliography}

\clearpage

\section*{Appendix A: Posterior computations}
Let $s_i=(s_{i,y},s_{i,x})$ denote cluster membership for subject $i$. Note that the value of $s_{i,x}$ is only meaningful in conjunction with $s_{i,y}$, as it describes which cluster within $s_{i,y}$ it belongs. We extend Algorithm 8 of \citet{neal:2000} to accommodate the nested clustering.  The basic steps in the Gibbs sampler are as follows. 
We sample $s_i$ for each subject, and then, given $s$, we sample parameters $\theta$ and $\omega$ from their conditional distributions, given data. Denote by $\theta^*_j$ the $\theta$ that is associated with the $j$th of the currently non-empty clusters. $\omega^*_{l|j}$ is defined similarly.

Denote by $k$ the current number of $y$-clusters (i.e., the number of unique values of $s_y$) and by $k_j$ the number of $x$-subclusters of the $j$th $y$-cluster. We use $-i$ notation to indicate that the $i$th value is removed (e.g., $y^{-i}$ is the vector of outcomes excluding subject $i$).

For the step involving drawing cluster membership, each subject can be assigned to one of the non-empty clusters, to one of $m$ new $y$-clusters, or to one of $m$ new $x$-clusters within each existing $y$-cluster. Each possible new cluster is associated with auxiliary parameters. In the simulations and data analysis, we used $m=5$.

\noindent {\bf{Update cluster membership $s_i$}}

First, we need to do some relabeling and generate some auxiliary parameters for potential new clusters, before we can update $s_i$. For subject $i$, denote by $k^{-i}$ the number of unique $y$-clusters currently non-empty if you exclude subject $i$, and $k_{j}^{-i}$ the number of unique $x$-subclusters of the $j$th $y$-cluster that are non-empty if you exclude subject $i$. Label these clusters that are currently occupied by the subjects other than $i$ $\{1,\cdots,k^{-i}\}$ for the $y$-clusters and $\{1,\cdots,k_{j}^{-i}\}$ for the $x$-subcluster of the  $j$th $y$-cluster.   Denote by $n_j^{-i}$ and $n_{l|j}^{-i}$ the number of subjects in the corresponding subclusters, excluding subject $i$. If the current value of $s_i$ is in one of these clusters, then draw values of $\theta^*$ and $\omega^*$ from prior distributions $P_{0\theta}$ and $P_{0\omega}$ for $y$-clusters $\{k^{-i}+1,\cdots,k^{-i}+m\}$. If the current value of $s_i$ is in an existing $y$-cluster (say, $y$-cluster $j$) but not an existing $x$-subcluster, then set its current cluster to $s_i=(j,k_j^{-i}+1)$ and draw $m-1$ values of $\omega^*$ from its prior distribution and assign them to $\{k^{-i}_j+2,\cdots,k^{-i}_j+m\}$. In addition, draw $m$ sets of parameters from the priors for the other clusters and subclusters. Finally, if the current value of $s_{yi}$ does not correspond with any of the $k^{-i}$ $y$-clusters, then $s_{yi}=k^{-i}+1$ and then draw $\theta^*$ and $\omega^*$ for $y$-clusters $\{k^{-i}+2,\cdots,k^{-i}+m\}$ and $x$-subclusters $\{k^{-i}_j+1,\cdots,k^{-i}_j+m\}$ for $j=1,\cdots,k^{-i}$. At this point, all of the occupied and extra clusters have $\theta^*$ and $\omega^*$ parameters associated with it. We can now draw a new value of $s_i$.

Draw a new value $s_i$ as follows. $P(s_i=(j,l)|s_{-i},\theta^*,\omega^*,x,y)=$

\[
=\left\{
\begin{array}{ll}
b\frac{n_j^{-i}n_{l|j}^{-i}}{n_j^{-i}+\alpha_\omega}K(y_i|x_i,\theta_j^*)K(x_i|\omega_{l|j}^*), &  {\rm{for}} \ 1\le j\le k_j^{-i} \ {\rm{and}} \ 1\le l\le k_{l|j}^{-i} \\
b\frac{n_j^{-i}\alpha_\omega/m}{n_j^{-i}+\alpha_\omega}K(y_i|x_i,\theta_j^*)K(x_i|\omega_{l|j}^*), &  {\rm{for}} \ 1\le j\le k_j^{-i} \ {\rm{and}} \ k_{l|j}^{-i}< l\le k_{l|j}^{-i}+m \\
b\frac{\alpha_\theta}{m}K(y_i|x_i,\theta_j^*)K(x_i|\omega_{l|j}^*), &  {\rm{for}} \ k_j^{-i}< j\le k_j^{-i}+m 
\end{array}
\right.
\]
where $b$ is a constant such that $1=\sum_{j=1}^{k^{-i}+m} \sum_{l=1}^{k_j^{-i}+m} P(s_i=(j,l)|s_{-i},\theta^*,\omega^*,x,y)$.

This step is done for each $i=1,\cdots,n$.

\noindent {\bf{Update parameters $\theta^*$ and $\omega^*$}}

For each unique  $j$ in $s_y=\{s_{y1},\cdots,s_{yn}\}$, update $\theta^*_j$ from
\[
p(\theta_j^*|s,y,x,\theta^*_{-j},\omega^*)\propto p_{0\theta}(\theta^*_j)\prod_{i: s_{yi}=j} K(y_i|x_i,\theta_j^*).
\]
This update will be a standard update from a Bayesian regression.

For each unique $(j,l)$ in $s=\{s_1,\cdots,s_n\}$, update $\omega^*_{l|j}$ from
\[
p(\omega_{l|j}^*|s,y,x,\theta^*,\omega^*_{-l|j})\propto p_{0\omega}(\omega^*_{l|j})\prod_{i: s_i=(j,l)} K(x_i|\omega_{l|j}^*).
\]
Consider the situation where the first $p_1$ variables in $L$ are binary, and the remaining $p_2$ variables are continuous. We assume 
\[
p(x_{i,r}|\omega_i)=\rm{Bern}(\pi_i^r)\]
 with 
\[p_{0\omega}(\pi_i^r)= Beta(a_{x},b_{x}),  \ r=1,\cdots,q-1+p_1.\]
Thus, we update $\pi_{l|j}^{r*}$ from Beta$\left(a_{x}+\sum_{i:s_i=(j,l)} x_{i,r},b_{x}+n_{l|j}-\sum_{i:s_i=(j,l)}x_{i,r}\right)$. 

For the last $p_2$ $x$ variables, we assume 
\[p(x_{i,r}|\omega_i)=N(\mu_i^r,\tau^{2,r}_i)\]
 with 
\begin{align*}
p_{0\omega}(\mu_i^r|\tau^{2,r}_i)&=N(\mu_0,\tau^{2,r}_i/c_0), \\
p_{0\omega}(\tau^{2,r}_i)&= Inv-\chi^2(\nu_0,\tau_0^2). 
\end{align*}
We can then update  $\mu_{l|j}^{r*}$ and $\tau_{l|j}^{2,r*}$ from normal and scale inv-$\chi^2$ distributions.
\[
\tau_{l|j}^{2,r*}|rest \sim Inv-\chi^2\left(\nu_0+n_{l|j}, \frac{\nu_0\tau_0^2+(n_{l|j}-1)s^{2,r}_{l|j} +\frac{c_0n_{l|j}}{c_0+n_{l|j}}(\overline{x}_{l|j}^r-\mu_0)^2}{\nu_0+n_{l|j}}\right)
\]
\[
\mu_{l|j}^{r*}|rest \sim N\left(\frac{\frac{c_0}{\tau_{l|j}^{2,*}}\mu_0+\frac{n_{l|j}}{\tau_{l|j}^{2,*}}\overline{x}_{l|j}^r}{\frac{c_0}{\tau_{l|j}^{2,*}}+\frac{n_{l|j}}{\tau_{l|j}^{2,*}}} , \frac{1}{\frac{c_0}{\tau_{l|j}^{2,*}}+\frac{n_{l|j}}{\tau_{l|j}^{2,*}}} \right)
\]
where $\overline{x}_{l|j}^r$ and $s^{2,r}_{l|j}$ are the sample mean and sample standard deviation, respectively, of the $r$th covariate among subjects with $s=(j,l)$.

\noindent {\bf{Update hyperparameters}}

Update $\alpha_\theta$ and $\alpha_\omega(\theta)$. We specify Gam$(a_0,b_0)$. To update $\alpha_\theta$
first draw $\eta\sim Beta(\alpha_\theta+1, n)$. Next, set draw $\alpha_\theta$ from 
\[
\pi Gam(a_0+k, b_0-\log(\eta))+(1-\pi)Gam(a_0+k, b_0-\log(\eta))
\]
where $\pi=\frac{\frac{k}{n(1-\log(\eta))}}{1+\frac{k}{n(1-\log(\eta))}}$ \citep{esco:west:1995}. To update $\alpha_\omega$, we use Metropolis-Hastings, where 
\[
p(\alpha_\omega|rest)\propto p(\alpha_\omega)\alpha^{\sum_{j=1}^k (k_j-1)}\prod_{j=1}^k (\alpha_\omega+n_j)\beta(\alpha_\omega+1,n_j).
\]

\clearpage

\section*{Appendix B: Additional data analysis results} 
Table S1 shows the use of mtNRTI-containing ART regimens vs other NRTI ART regimens as first-line therapy from 2002-2009. The use of mtNRTI ART regimens began as the majority and declined over time while the use of other NRTI ART regimens started low and increased. Figure S1 shows the trace plot and posterior distribution for the average causal relative risk of death comparing use of mtNRTI vs other NRTI ART regimens in the HIV/HCV cohort study example.\\

 \renewcommand\thetable{S1}
\begin{table}[h]
\caption{Number of subjects newly initiating an ART regimen that includes an mtNRTI or other NRTI, between the years 2002 and 2009.}
\begin{tabular}{lcccccccc}
\hline
\hline
Exposure & 2002 & 2003 & 2004 & 2005 & 2006 & 2007 & 2008 & 2009 \\
\hline
mtNRTI  & 250  &233 & 163&   89&   42&   23&   24&   12\\
Other NRTI &15 &  37 &  63 & 145  &158 & 161 & 180 & 152 \\
\hline
\end{tabular}
\end{table}

\renewcommand\thefigure{S1}
\begin{figure}[h]
\begin{center}
 \includegraphics[width=5in]{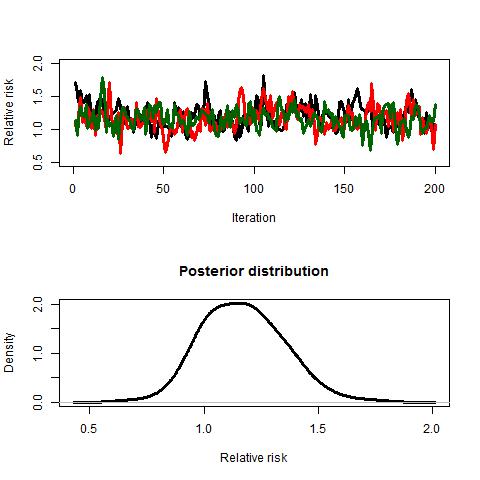}
\caption{Trace plot (top) and posterior density plot (bottom) of the causal effect in the HIV/HCV cohort study example.}  
\label{fig:trace}
\end{center}
\end{figure}

\end{document}